\documentclass[preprint,aps,pra]{revtex4-2}

\usepackage{siunitx}
\DeclareSIUnit{\gauss}{G}
\usepackage{graphicx}

\usepackage[colorlinks=false,  % must be last 
pdfborder={0 0 2}      
]{hyperref}         % hyperlinking

\newcommand{\autorefs}[1]{\hyperref[#1]{Figs.~\ref*{#1}}}
\newcommand{\Autoref}[1]{\hyperref[#1]{Figure~\ref*{#1}}}
\newcommand{\Autorefs}[1]{\hyperref[#1]{Figures~\ref*{#1}}}

\usepackage{cleveref}  % for \autoref{}, \cref{}

\begin{document}
% [solution] + [problem]
% [bifurcation of atoms is washed out in strong magnetic field (strong interactions)] + [what happens to Li-6 in a shaken optical lattice with tunable interaction]
\title{Interaction and confinement effects for a \texorpdfstring{${}^6\mathrm{Li}_2$}{6Li2} condensate in a shaken lattice}

% APS
%\title{Band populations and interaction effects of a ${}^6\mathrm{Li}_2$ molecular condensate in a shaken optical lattice}

\author{Yang-Hao Lee}
\author{Duc Trong Le}
\author{John Karnosh}
\author{Colin V. Parker}
\affiliation{Georgia Institute of Technology, Atlanta, GA USA}

\begin{abstract}
A shaken optical lattice can form Floquet-engineered band structures. For a one-dimensional lattice, this method can hybridize the ground and first excited bands into a double-well dispersion. A molecular Bose-Einstein condensate (mBEC) loaded into this band bifurcates into two clusters at opposite quasimomenta, but short cluster lifetimes (\qtyrange{2}{3}{\milli\second}) have limited earlier work. We improve our $^6$Li mBEC experiment by aligning the lattice axis with the trap and using a multifrequency shaking sequence to counterdiabatically load into the double-well minima. Together, these improvements allow us to extend the experiment timescale beyond \qty{10}{\milli\second}. At these longer times the momentum distribution depends qualitatively on the confinement. With stronger confinement in the lattice direction, there is a peak in the momentum distribution at the Brillouin zone edge. Shaking-frequency spectroscopy confirms this zone edge population and shows it splits roughly equally between the ground and first excited bands. The longer lifetime also lets us ramp the magnetic field from \qty{650}{\gauss}, where molecules interact weakly, to \qty{800}{\gauss}, near the Feshbach resonance. As the interaction strengthens, the bifurcation weakens and disappears. We suggest that the increased molecular size causes this change.

\end{abstract}
\maketitle

\section{Introduction}
%--------- first introduce about shaken optical lattice -------------
Shaking an optical lattice adds couplings and reshapes the band structure of the atoms it holds. This technique, known as Floquet band engineering \cite{Holthaus2015FloquetEngineering}, is widely applied in ultracold quantum gas experiments to simulate a variety of novel quantum systems beyond static lattices \cite{Eckardt2017ColloquiumAtomic, Goldman2014PeriodicallyDriven, Weitenberg2021TailoringQuantum, Dotti2025MeasuringLocalization}. Such driven systems can also be interpreted as lattices in momentum space, with each drive frequency component representing a tunneling amplitude \cite{Gadway2015AtomopticsApproach}. Shaking a one-dimensional optical lattice near the resonance between the ground and first excited bands hybridizes the two bands into a double-well dispersion. The transition from a single ground band minimum to two minima at opposite quasimomentum with an unstable Brillouin zone center is analogous to ferromagnetic ordering \cite{Parker2013DirectObservation}. When the shaking is instead strong and off-resonant, it results in an inverted band with negative effective mass\cite{Lignier2007DynamicalControl}. Condensates in these engineered bands have been used to study formation of domains \cite{Parker2013DirectObservation, Liu2016ExoticDomain, Yao2022DomainwallDynamics}, roton-maxon excitations \cite{Ha2015RotonmaxonExcitation}, and Floquet solitons with negative effective mass \cite{Mitchell2021FloquetSolitons}.

% ----------- more on lithium, interaction, scattering length... --------
% main sentence:
% ideas: fermion, wide tunability, feshbach resonance, interesting for shaken lattice experiments
Interactions shape these effects. They set the characteristics of the transition \cite{Zheng2014StrongInteractiona, Song2022RealizingDiscontinuous}, and open collision channels that heat the cloud \cite{Choudhury2014StabilityFloquet, Choudhury2015StabilityBoseEinstein}. For fermionic ${}^6\mathrm{Li}$, a broad magnetic Feshbach resonance near \qty{832}{\gauss} tunes interactions over a wide range \cite{Chin2010FeshbachResonances, Zurn2013PreciseCharacterization}. On its BEC side, atoms pair into weakly bound molecules that form a molecular Bose-Einstein condensate (mBEC) \cite{Jochim2003BoseEinsteinCondensation, Zwierlein2003ObservationBoseEinstein}, with the molecule--molecule scattering length growing by more than one order of magnitude when approaching the resonance \cite{Bartenstein2005PreciseDetermination}. On the other side of the resonance, the mBEC turns continuously into a BCS superfluid \cite{Giorgini2008TheoryUltracold, Chen2024WhenSuperconductivity}. An ultracold ${}^6\mathrm{Li}$ gas in a shaken lattice could therefore realize the BEC-BCS crossover in an engineered band, where theory predicts exotic pairing such as Fulde--Ferrell--Larkin--Ovchinnikov (FFLO) phases \cite{Zheng2015FloquetFuldeFerrellLarkinOvchinnikov, Zheng2016FuldeFerrellSuperfluids}. Interactions have already been tuned in a static lattice \cite{Kiefer2023UltracoldFeshbach} and in driven Hubbard lattices \cite{Messer2018FloquetDynamics, Desbuquois2017ControllingFloquet}. In this article, we explore the effects of interactions on ${}^6\mathrm{Li}_2$ mBEC in a shaken optical lattice.

%However, no research (checkingTODO) has tuned the interactions for a condensate in a shaken-lattice double-well band. We therefore ask what happens to such a state when the interactions are tuned.

% ----------- previous work ----------
Exploring these effects requires an mBEC that lives long enough in a double-well band. Slow ramp-up of the shaking forms domains \cite{Parker2013DirectObservation, Clark2016UniversalSpacetime}, but the cloud would heat up during the ramp with strong interactions \cite{Lellouch2017ParametricInstability, Wintersperger2020ParametricInstabilities}. 
In our earlier work \cite{Wang2023InstabilityMomentum}, a ${}^6\mathrm{Li}_2$ mBEC was instead exposed to lattice shaking with rapid ramp-up and shaken within \qtyrange{2}{3}{\milli\second} in a shallow lattice. Leaving the mBEC at a sharp maximum of the dispersion, the transition resulted in the mBEC bifurcating cleanly into two clusters at opposite quasimomenta. However, these clusters collided and heated when evolving in the band and only stayed distinguishable for \qty{2}{\milli\second}, too short to tune the field afterward.

% ----------- what we've done ----------
In this work, we realize a soft landing of the bifurcated clusters, extending cluster lifetime to beyond \qty{10}{\milli\second}. First, a dipole trap perpendicular to the lattice confines the dynamics to the lattice direction and suppresses the transverse motion from our earlier work. Second, a multifrequency shaking sequence counterdiabatically loads the shaken molecules in the double-well minima, removing excess potential energy from the trap. The extra lifetime lets us equilibrate samples for far longer, and ramp the magnetic field after the sample is stabilized.

An interesting consequence we observe is that a deeper perpendicular trap (i.e. tighter confinement along the direction of the lattice itself) creates a peak in the momentum distribution at the Brillouin zone edge.
Using shaking-frequency spectroscopy \cite{Heinze2011MultibandSpectroscopy}, we show that this population splits roughly equally between the ground and first excited bands. We then ramp the magnetic field from \qty{650}{\gauss} toward the Feshbach resonance and find that the bifurcation weakens as the interactions strengthen. A comparison of two field sequences shows that heating cannot explain this loss. We therefore suggest that strong interactions around the Feshbach resonance wash out the bifurcation after it forms. The tunable interactions add a control parameter to Floquet-engineered bands, and open the way to study interaction-driven instabilities and strongly interacting pairs. These results are a first step toward realizing strongly interacting ${}^6\mathrm{Li}$ pairs and ultimately the BEC-BCS crossover in exotic bands.

% \section{Literature Review}
% \input{2-lilterature review.tex}

\section{Experimental setup}

An ultracold ${}^{6}\mathrm{Li}_2$ mBEC is prepared in a \qty{1070}{\nano\meter} crossed optical dipole trap through evaporative cooling as in previous work \cite{Wang2023InstabilityMomentum}. The mBEC is transferred to a separate, \qty{1064}{\nano\meter} perpendicular crossed dipole trap generated by a single-frequency laser, which consists of a retro-reflected lattice beam and a perpendicular trapping beam (which is not reflected). We define the direction of the incoming lattice beam as the $z$ axis, the direction of the trapping beam as the $y$ axis, and the direction of imaging projection and gravity as the $x$ axis. During the transfer, the \qty{1070}{\nano\meter} crossed dipole trap, whose beams make an angle of \qty{59}{\degree} with the $z$ axis, is gradually ramped off to leave the potential symmetric in the positive and negative $z$ direction.

After the molecules are loaded into the perpendicular crossed dipole trap, the returning lattice beam is ramped up to form a one dimensional optical lattice with lattice depth $V_0 \approx 1.9\,E_\mathrm{R}$, where $E_\mathrm{R} = \hbar^2k_\mathrm{L}^2/(2m)$ is the recoil energy, $k_\mathrm{L} = \pi/a$ is the lattice momentum, and $a$ is the lattice spacing of \qty{532}{\nano\meter}. Two acousto-optic modulators (AOMs) diffract the lattice beam and modulate the returning beam. The input signal to one of the AOMs is modulated by an in-phase and quadrature modulator, as described in previous work~\cite{Wang2023InstabilityMomentum}. We hold the in-phase component fixed and modulate quadrature component periodically at the shaking frequency $f_\mathrm{shk}$. This modulation shifts the phase returning beam and therefore displaces the lattice sites along the $z$ axis. We quantify the shaking by the dimensionless peak-to-peak shaking amplitude $\theta_{max}/2\pi$, the peak-to-peak lattice displacement in units of the lattice spacing. The in-phase amplitude of \qty{0.3}{V} and the quadrature peak-to-peak amplitude of \qty{0.330}{V} give a shaking amplitude $\theta_{max}/2\pi \approx 0.16$. The lattice holds about \num{22000} molecules with a condensate fraction of 0.41, obtained from a bimodal fitting. The temperature is estimated at about \qty{160}{\nano\kelvin} from the thermal momentum width after \qty{7}{\milli\second} time-of-flight (TOF).

% $f_\mathrm{shk}$ is chosen to nearly match the splitting between the ground band and any of the three lowest excited bands at various points in the Brillouin zone.

%For most experiments, the optical lattice is first shaken at \qty{60}{\kilo\hertz} for \qty{0.2}{\milli\second}, matching the splitting between the two lowest undressed bands near the zone center. This forms a double-well band structure in the lowest dressed band and allows atoms to quickly evolve into the opposite momentum minima of this band.

% Subsequently, $f_\mathrm{shk}$ is ramped down to \qty{54}{\kilo\hertz} for the rest of the shaking time $t_\mathrm{shk}$. This increased the separation between dressed band minima and suppresses the oscillation of the bifurcated clusters, which extends the lifetime of atoms in the shaken optical lattice up to \qty{12}{\milli\second}.

\begin{figure}[ht]
\centering
\includegraphics[width=1\linewidth]{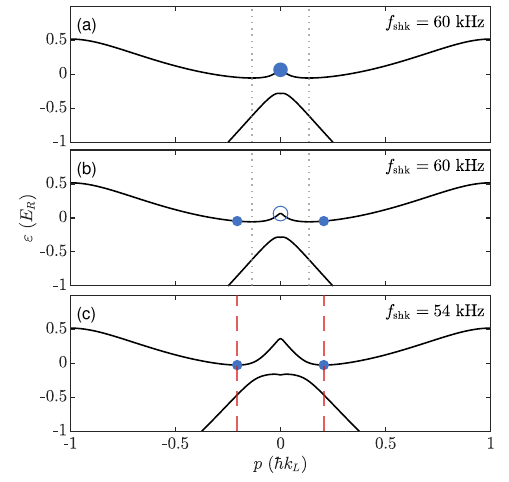}
\caption{Schematic of the counterdiabatic loading process. (a) At the start of the shaking, a condensate is located at the unstable zone center of the dispersion. (b) The cluster bifurcates and evolves pass the two double-well minimum through exchanging kinetic energy with potential. (c) $f_\mathrm{shk}$ is ramped to a lower value. The quasimomentum splitting between the two minima in the dispersion becomes wider and catches the clusters close to the new double-well minimum.}
\label{fig:softLanding_demo}
\end{figure}

For most experiments, the shaking frequency is set to \qty{60}{\kilo\hertz} at the start of the shaking. Initial instability around the zone center causes the atoms to fall into states with small $+z$ or $-z$ momenta. In the negative curvature region of the dispersion, the effective mass is negative. Therefore, the direction of group velocity is opposite to the direction of the crystal momentum. Since the initial group velocity is non-zero, the atoms in real space move away from the center of the dipole trap. The magnitude of crystal momentum also increases in the direction opposite to the trapping potential gradient, and the atoms gradually evolve toward the lower energy states at the double-well minima while gaining displacement in the opposite $z$. As the sign of effective mass flips when entering into the positive curvature region, the quasimomentum increase starts to reduce. After the states move across the double-well minima, the atoms start moving back towards the dipole trap center. To remove potential energy from the system, $f_\mathrm{shk}$ is ramped down to \qty{54}{\kilo\hertz} in \qty{0.2}{\milli\second} after evolving at $f_\mathrm{shk} = \qty{60}{\kilo\hertz}$ for \qty{0.2}{\milli\second}. The quasimomentum splitting between the two minima in the dispersion becomes wider to match the quasimomenta of the states. This reduces the group velocity of atoms as they approach the dipole trap center again, taking energy out of the system. The oscillation around the double-well minima is suppressed and this leads to less atoms loss. The bifurcated clusters therefore survive to $t_\mathrm{shk} = \qty{12}{\milli\second}$ in a shaken optical lattice. A schematic of this process is illustrated in \autoref{fig:softLanding_demo}.

After a shaking sequence is finished, the atoms are released into free space and projected to plane wave states, whose different momenta can be observed by TOF imaging. The typical trapping frequencies in $x$, $y$, and $z$ are \qtylist{204;47;203}{\hertz}, respectively. A weaker confinement by the trapping beam perpendicular to the lattice can also be used, in which case the trapping
frequency in $z$ reduces to \qty{102}{\hertz}. Since the confinement along $x$ is primarily from this same beam, the $x$ direction trapping will be reduced proportionally. Varying the confinement along the lattice direction affects the equilibrium momentum distribution significantly.

\section{Results}
% \input{text-exp result.tex}
% Figure 1.\\
% Two stage shaking \\
% bifurcated state survives for over 10 ms\\
% Zone edge signal when varying external harmonic trap (perpendicular one) strength\\

\begin{figure}[ht]
\centering
\includegraphics[width=1\linewidth]{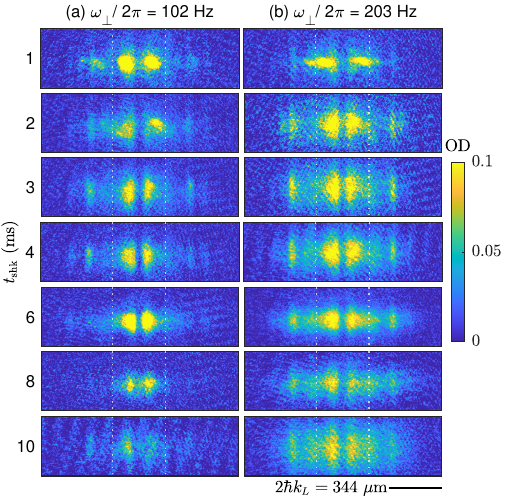}
\caption{Evolution of the momentum distribution (lattice direction is horizontal) as a function of shaking time $t_{\mathrm{shk}}$ under two trapping conditions. Trapping frequencies along the lattice axis are (a) \qty{102}{\hertz} and (b) \qty{203}{\hertz}. White dashed lines mark the first Brillouin zone edge.}
\label{fig:tshk_evol}
\end{figure}

The $t_{\mathrm{shk}}$ evolution (imaged after \qty{7}{\milli\second} TOF) with trapping frequencies along $z$ being \qtylist{102;203}{\hertz} are shown in \autoref{fig:tshk_evol}. Unlike in previous experiments, the lattice is along a principal axis of the trap, and the resulting dynamics are primarily along $z$. The horizontal axis in these images is $z$ axis displacement after the TOF, and can be understood as measuring the $z$ axis momentum distribution. With a higher trapping frequency, a maximum of signal located around the Brillouin zone edge is noticeable, as is more population near the center of the second zone. The latter population can be attributed to atoms near the center of the zone in the lowest dressed band, as dressed states near the center are composed primarily of bare states with $\pm 1$ unit of lattice momentum, while those near and outside the dressed band minima are composed primarily of bare states with zero momentum.

% Figure 2.
% Tried to see where the zone edge signal comes from through in-situ imaging \\
% \begin{figure}[ht]
% \centering
% \includegraphics[width=0.9\linewidth]{fig2.pdf}
% \caption{test figure 2}
% \label{fig:TraceFit}
% \end{figure}

% To investigate the origin of the find where the side peaks come from, we imaged the cloud after a short time of flight, from \qty{0.02}{\milli\second}, essentially in situ, to \qty{3}{\milli\second}. To amplify the side-peak signal, we stopped the shaking just before release and shifted the frequency of one returning lattice beam with its AOM. This shift drives a Bragg transition, which transfers a molecule by one two-photon recoil ($2\hbar k_\mathrm{L}$) when the energy difference between the initial and final momentum states matches the frequency shift. Here, the frequency shift is tuned to the transition from each bifurcated peak to its neighboring momentum order, so a fraction of each bifurcated peak moves into a first-order peak. We then subtracted the image taken with a pulse toward $-z$ from the image taken with a pulse toward $+z$. This makes the weak first-order peaks stand out while the bifurcated peaks remain visible as lobes of opposite sign [Figure  \autoref{fig:TraceFit}(a)--\autoref{fig:TraceFit}(c)]. Each first-order peak keeps a fixed momentum offset of $2\hbar k_\mathrm{L}$ from its zeroth-order partner, so one trajectory model describes all four peaks. The differential images therefore constrain the trajectories far more tightly than the raw TOF images do.

It is not surprising to see the momentum distribution increase in width as the confinement increases. This would be predicted by basic uncertainty principle arguments. It is also true that higher density should lead to a higher momentum spread in ground state occupancy in a Bogoliubov picture. However, this qualitatively fails to explain a local maximum in the momentum distribution near the zone edge. We propose two explanations. First, the maximum exists because atoms are promoted to the second dressed band, which has a local minimum at the zone edge, into which atoms might relax. Second, the maximum exists because interference between the two primary peaks creates a standing wave, and the resulting density modulation is strong enough to scatter atoms into higher momentum states. For example, two condensates at momenta $-\hbar k^*$ and $\hbar k^*$ produce a standing wave with wavevector $2k^*$, which can scatter and create population with momenta $-3\hbar k^*$ and $3\hbar k^*$. Note that in the second case, the observation of significant interference suggests conditions different from true bosonic shaken lattice experiments, in which domain formation limits the formation of interference \cite{Parker2013DirectObservation, Clark2016UniversalSpacetime, Liu2016ExoticDomain, Feng2018CoherentInflationary, Yao2022DomainwallDynamics}.

% original:
% Distinguishing these possibilities is challenging, because the split condensate survives best in conditions where $3k^*$ nearly matches the zone edge. We can, however, attempt to determine whether the signal comes from occupation in the ground band (index 0) or first excited band (index 1), using shaking-frequency spectroscopy. A sample is prepared using the normal two-stage shaking sequence to shake for $t_\mathrm{shk} = \qty{4.9}{\milli\second}$. Next, the shaking frequency is ramped to a new value which can match excitations from bands 0 or 1 into the next two highest bands (with indices 2 and 3), and we shake for another \qty{0.1}{\milli\second} before releasing to free space. This couples the bands and allows some population to transfer to the higher bands. The higher Bloch bands are superpositions of plane wave states with momentum differing by an integer multiple of the lattice momentum. When abruptly releasing to free space, we see depletion of the $p=\hbar q$ diffraction peak accompanied by increased population in the $p=\hbar(q\pm2k_\mathrm{L})$ peaks. 
\begin{figure}[ht]
\centering
\includegraphics[width=0.8\linewidth]{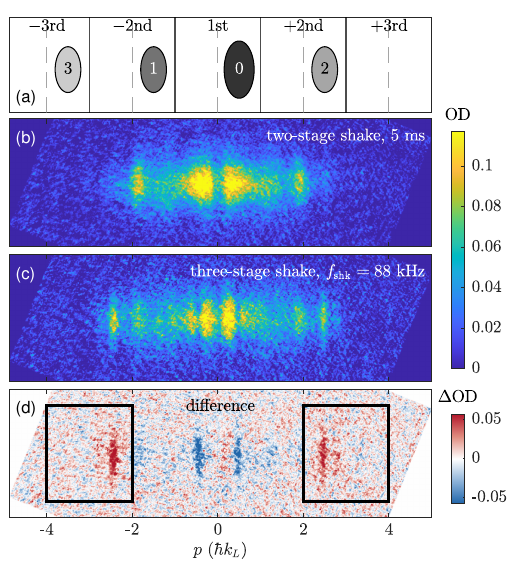}
\caption{
(a) Schematic of the main plane-wave component of a population in bands 0, 1, 2, and 3 at some $p \ge 0$ in the first Brillouin zone, in the shallow-lattice limit. Solid lines separate the zones, each with width $2\hbar k_\mathrm{L}$, and dashed lines mark the lattice centers. (b) An absorption image of a standard two-stage shaking for $t_\mathrm{shk} = \qty{5}{\milli\second}$ as a baseline reference. (c) An absorption image of two-stage shaking for \qty{4.9}{\milli\second}, then \qty{0.1}{\milli\second} at $f_\mathrm{shk} = \qty{88}{\kilo\hertz}$. The $p = \hbar q$ peak is depleted, and the $p = \hbar(q \pm 2k_\mathrm{L})$ peaks gain population. (d) Differential image of (c) minus (b). The signal in the two boxes are averaged together, with the left box mirrored, and summed vertically to generate a row of \autoref{fig:shkFreqSpec_scan}.
%Appen figure: shake frequency spectroscopy demo (explaining \autoref{fig:shkFreqSpec_scan})
}
\label{fig:shkFreqSpec_demo}
\end{figure}
% rewrite
Distinguishing these possibilities is challenging, because the split condensate survives best in conditions where $3k^*$ nearly matches the zone edge. We can, however, attempt to determine whether the signal comes from occupation in the ground band (index 0) or first excited band (index 1), using shaking-frequency spectroscopy. 
\Autoref{fig:shkFreqSpec_demo} illustrates the spectroscopy process. To identify the transferred population, a standard two-stage shaking for $t_\mathrm{shk} = \qty{5}{\milli\second}$ is used as a baseline. A second sequence shakes for \qty{4.9}{\milli\second}, after which the shaking frequency is ramped to a new value which can match excitations from bands 0 or 1 into the next two highest bands, and the atoms are shaken for another \qty{0.1}{\milli\second} before releasing to free space. This couples the bands and allows some population to transfer to the higher bands. When abruptly releasing to free space, we see depletion of the $p=\hbar q$ diffraction peak accompanied by increased population in the $p=\hbar(q\pm2k_\mathrm{L})$ peaks. A differential image is obtained by subtracting these two. This leaves information about how much signal in the window $p = 0$--$2 \hbar k_\mathrm{L}$, which is the main range of plane waves that constructed from bands 0 and 1, is transferred into the window $p = 2\hbar k_\mathrm{L}$--$4\hbar k_\mathrm{L}$, which is the main range of the plane waves that construct bands 2 and 3.

% original:
% \autoref{fig:shkFreqSpec_scan} shows the \qty{7}{\milli\second} TOF atom signal difference obtained by taking the three-stage shaking signal with the sequence described above (the third stage being the spectroscopy pulse) and subtracting a standard \qty{5}{\milli\second} two-stage shaking sequence background. The figure shows the mean value of population transferred from the window $p = 0-2 \hbar k_\mathrm{L}$, which is the main range of plane waves that construct bands 0 and 1, into the window $p = 2-4\hbar k_\mathrm{L}$, which is the main range of the plane waves that construct bands 2 and 3.
% rewrite:

\begin{figure}[ht]
\centering
\includegraphics[width=1\linewidth]{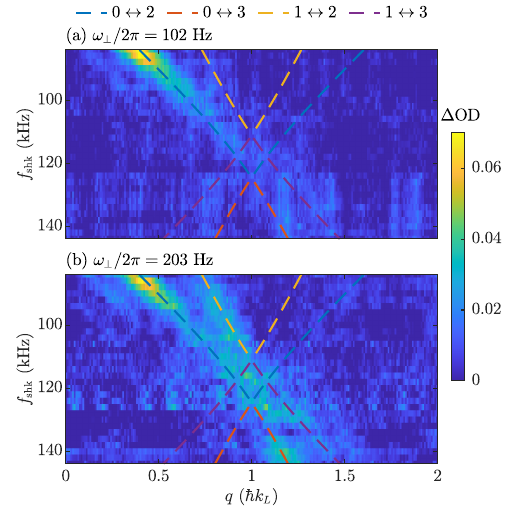}
\caption{Shaking-frequency spectroscopy scan from $f_\mathrm{shk} = 84$ to \qty{144}{\kilo\hertz}. Trapping frequencies along the lattice axis are (a)\qty{102}{\hertz} and (b) \qty{203}{\hertz}. The calculated band crossing points are overlaid on top as dashed lines. The colors of the dashed line indicates which two bands are hybridizing and more likely to have population transfer.}
\label{fig:shkFreqSpec_scan}
\end{figure}

A shaking-frequency spectroscopy scan from $f_\mathrm{shk} = 84$ to \qty{144}{\kilo\hertz} obtained using the process described above is shown in \autoref{fig:shkFreqSpec_scan}. At $\omega_\perp/2\pi = \qty{102}{\hertz}$ [\autoref{fig:shkFreqSpec_scan}(a)], the signal near the zone edge ($q = \hbar k_\mathrm{L}$) is weak, and only the first band is occupied, consistent with \autoref{fig:tshk_evol}(a). At $\omega_\perp/2\pi = \qty{203}{\hertz}$ [\autoref{fig:shkFreqSpec_scan}(b)], however, the zone edge carries a strong signal over shaking frequencies from \qty{100}{\kilo\hertz} to \qty{140}{\kilo\hertz}. To identify its origin, we calculate the quasimomenta at which the shaking frequency matches the energy difference between two bands, and overlay them on top as dashed lines. Near the zone edge, the crossings from band 0 into bands 2 and 3 lie near \qty{124}{\kilo\hertz}, whereas the crossings from band 1 lie near \qty{110}{\kilo\hertz}. The zone edge signal in \autoref{fig:shkFreqSpec_scan}(b) spans both ranges, without distinct peaks near either. This suggests that the non-interacting band description is no longer appropriate, and/or that both bands are nearly equally occupied. Neither case is consistent with relaxation to the minimum of band 1 causing the peak in the distribution near the zone edge.

\begin{figure}[ht]
\centering
\includegraphics[width=1\linewidth]{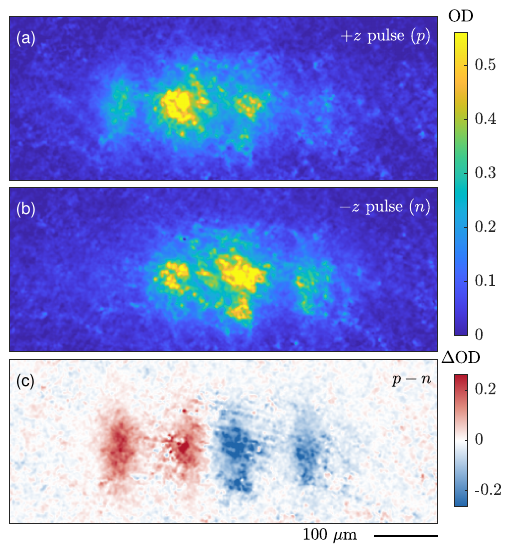}
\caption{Bragg differential imaging by transferring atoms into adjacent Brillouin zones. (a) transfer of the right bifurcated population in the first Brillouin zone to the left (b) transfer of the left bifurcated population in the first Brillouin zone to the right (c) Differential signal obtained by subtracting (a) by (b).}
\label{fig:bragg_demo}
\end{figure}

\begin{figure*}[!ht]
\centering
\includegraphics[width=0.95\textwidth]{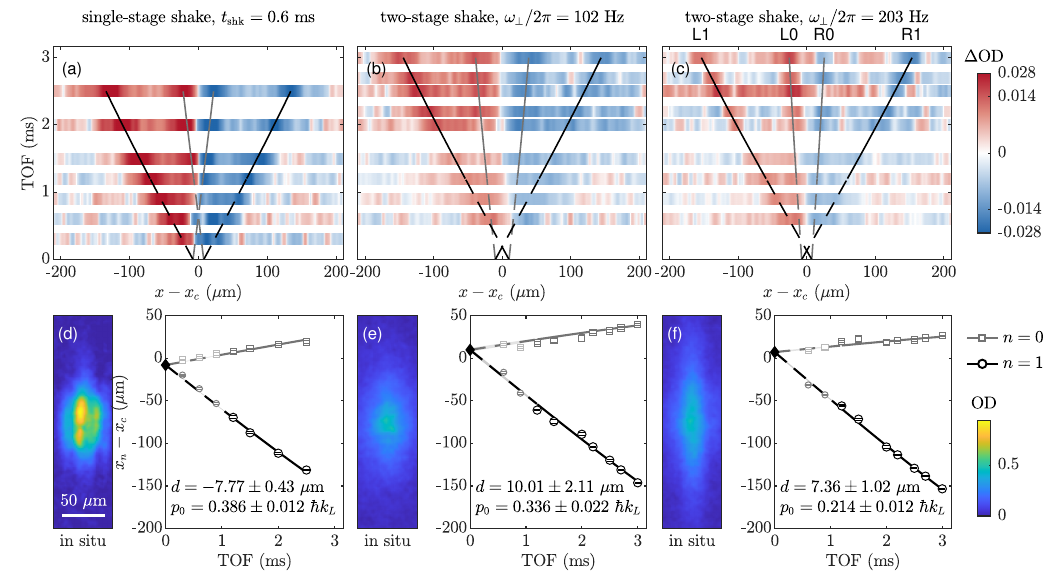}
\caption{Trajectory fitting of Bragg differential images similar to \autoref{fig:bragg_demo}. From left to right: single-stage shake with $t_\mathrm{shk}=$\qty{0.6}{\milli\second} at $f_\mathrm{shk} = \qty{60}{\kilo\hertz}$, and two-stage shakes with $\omega_\perp/2\pi = \qty{102}{\hertz}$ and \qty{203}{\hertz}. (a)--(c) Differential images $p - n$ (Bragg pulse toward $\pm z$) versus TOF, relative to the fitted center $x_c$, with a square-root color scale. Solid lines are fitted trajectories of the zeroth- (gray) and first-order (black) peaks, dashed where extrapolated below \qty{1}{\milli\second}. (d)--(f) In-situ optical density (left) and peak position versus TOF (right). Square and circle markers are single-TOF fits of the zeroth- and first-order peak positions, bands are jackknife variance $1\sigma$, and diamonds mark the fitted initial offset $d$.
}
\label{fig:TraceFit}
\end{figure*}
If we consider instead that the population near the zone edge is created by interactions, it is useful to investigate whether the sample forms domains and estimate the degree of overlap between them if so.
This is challenging, because the momentum distribution cannot be imaged at short times of flight, since the two clusters will overlap given their initial sizes, and the real space information is suppressed at long times of flight. However, by promoting population at low momentum into higher Brillouin zones, the separation can be made to appear earlier, on the order of a few \unit{\milli\second}. We image the cloud after a short time of flight, up to \qty{3}{\milli\second}, and amplify the second Brillouin zone signal using a Bragg transition. The Bragg transition is accomplished by removing the shaking, but changing the frequency of one of the AOMs in the retro-reflection path to create a moving lattice that transfers a molecule by two photon recoil ($2\hbar k_\mathrm{L}$) when the energy difference between the initial and final momentum states matches the frequency shift. Here, the frequency shift is tuned to the transition from each bifurcated peak to its neighboring momentum order, so a fraction of each bifurcated peak moves into a first-order peak. \Autoref{fig:bragg_demo} shows this transfers. To further amplify the signal, the image taken with a pulse toward $-z$ is subtracted from the image taken with a pulse toward $+z$ (i.e. with the lattice moving in the opposite direction). This makes the weak first-order peaks stand out while the bifurcated peaks remain visible as lobes of opposite sign [\autoref{fig:bragg_demo}(c)]. Each first-order peak keeps a fixed momentum offset of $2\hbar k_\mathrm{L}$ from its zeroth-order partner, so one trajectory model describes all four peaks. The Bragg differential images therefore constrain the trajectories far more tightly than the raw TOF images do.

Using this method, \autoref{fig:TraceFit} compares three shaking sequences. \Autoref{fig:TraceFit}(a) shows a single-stage shaking sequence with $t_\mathrm{shk} =$ \qty{0.6}{\milli\second}, which shakes with $f_\mathrm{shk}$ fixed at \qty{60}{\kilo\hertz} with $\omega_\perp/2\pi =$ \qty{203}{\hertz}. \Autorefs{fig:TraceFit}(b) and \ref{fig:TraceFit}(c) show the standard two-stage shaking sequence ramping from \qty{60}{\kilo\hertz} to \qty{54}{\kilo\hertz}, with $t_\mathrm{shk} =$ \qty{5}{\milli\second} and $\omega_\perp/2\pi = $ \qty{102}{\hertz} and \qty{203}{\hertz} in the second stage, respectively. The bottom row [\autorefs{fig:TraceFit}(d)--\ref{fig:TraceFit}(f)] shows the fitted peak positions with solid lines marking the fitted range, and dashed lines extrapolating the fit to zero time. We fit only the experiments with TOF $\geq$ \qty{1}{\milli\second}, because the zeroth- and first-order peaks are hard to resolve at short TOF. 

The fit parameter $d$ is the initial distance of the positive-momentum zeroth-order peak from the cloud center. A $d$ close to zero and comparable to half of the initial width implies the peaks are highly overlapped rather than formed into domains. For single-stage shaking, $d$ is negative, so the zeroth-order bifurcated peaks pass across each other in real space during the time of flight. Furthermore, the in-situ profile for this condition already shows evidence for the real-space separation of the clusters, with a central domain of positive momentum atoms and a slightly smaller domain of negative momentum atoms to it's right (there may even be a weaker domain on the far left). The single-stage shaking conditions were chosen deliberately to show this separation as a control. For both two-stage shaking sequences, $d$ is positive, indicating the zeroth-order peaks move apart and the first-order peaks cross instead. This implies that if domains have formed, the domains lie on opposite sides from their initial configuration. Additionally, the domains must either overlap substantially, or contain complex spatial patterns that remain unresolved. The latter possibility is particularly possible if the domains overlap along the imaging direction, since $x$ axis position is projected out and lost.

% \begin{figure*}[!ht]
% \centering
% \includegraphics[width=0.95\textwidth]{fig2b_trap_tmin1.0.pdf}
% \caption{test figure 2}
% \label{fig:TraceFit}
% \end{figure*}

% Figure 4.\\
% Show evolution with increasing field.\\
% increase interaction reduces the bifurcation\\

\begin{figure*}[ht]
\centering
\includegraphics[width=1\linewidth]{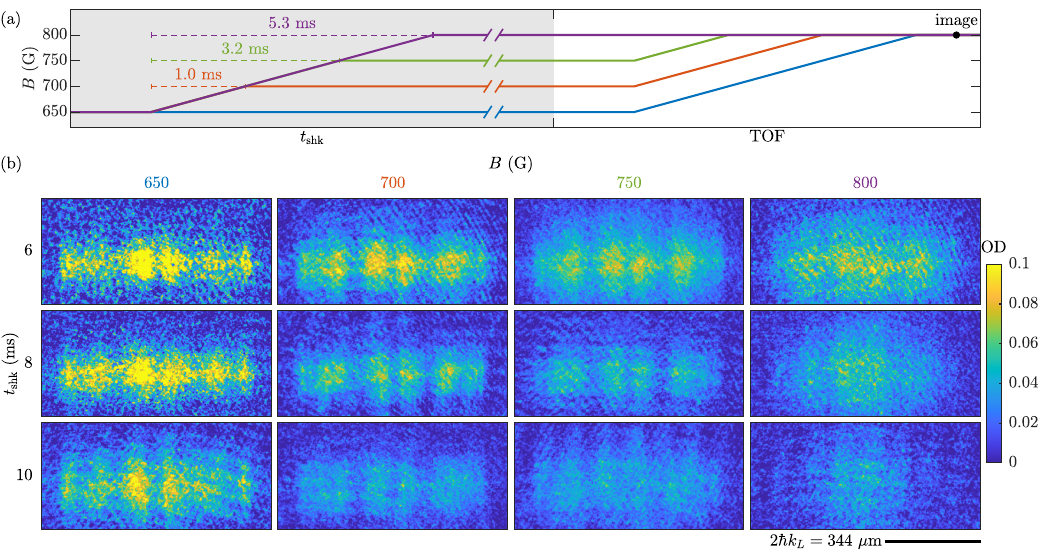}
\caption{Momentum profile with increasing interaction strength. (a) Field sequence (schematic, not to scale). The condensate is shaken for a total time $t_\mathrm{shk}$ (gray shading), starting at \qty{650}{\gauss}, and including a field ramp (see text for details). The break marks here indicate that the condensate is held for $t_\mathrm{shk}$ minus the ramp time and therefore varies between the rows in (b). During the \qty{7}{\milli\second} TOF, the field ramps to \qty{800}{\gauss}, and the image is taken at the indicated dot. (b) Absorption images of mBECs for $t_\mathrm{shk} =$ \qtylist{6;8;10}{\milli\second} (rows) and $B = $ \qtylist{650;700;750;800}{\gauss} (columns, colored as in (a)).}
\label{fig:TOF_evo_Bramp}
\end{figure*}

The longer cluster lifetime also lets us change the interaction strength after the bifurcation forms. To tune the interactions, we use the broad Feshbach resonance of ${}^{6}\mathrm{Li}$ near \qty{832}{\gauss}. Below the resonance, the atom-atom scattering length $a_\mathrm{s}$ grows with the magnetic field $B$, and the molecule--molecule scattering length follows as $a_{\mathrm{mm}} = 0.6\,a$ \cite{Petrov2004WeaklyBound}. Between 650 and \qty{800}{\gauss}, $a_{\mathrm{mm}}$ therefore increases from about $490\,a_0$ to $6500\,a_0$, where $a_0$ is the Bohr radius \cite{Bartenstein2005PreciseDetermination}.

\Autoref{fig:TOF_evo_Bramp}(a) shows the sequence. We first shake the condensate at \qty{650}{\gauss} with a two-stage shaking described in the experimental setup. During the second stage, the perpendicular trap frequency also ramps from $\omega_\perp/2\pi = $ \qty{203}{\hertz} to \qty{181}{\hertz} to slightly reduce temperature while keeping the side peak features. At the end of the shaking frequency and trap frequency ramp, the field begins to rise toward \qty{800}{\gauss} and stops at a target value $B$, which it reaches after 1.0, 3.2, and \qty{5.3}{\milli\second} for $B =$ 700, 750, and \qty{800}{\gauss}, respectively. The shaking continues at $B$ until the total shaking time $t_{\mathrm{shk}}$, when we release the cloud. About \qty{3}{\milli\second} into the \qty{7}{\milli\second} time of flight, we ramp the field to \qty{800}{\gauss}, so that every image is taken at the same field.

\Autoref{fig:TOF_evo_Bramp}(b) shows the resulting momentum distributions for $t_{\mathrm{shk}} = 6$, 8, and \qty{10}{\milli\second}. At \qty{650}{\gauss}, the central peak stays split into two clusters up to $t_{\mathrm{shk}} =$\qty{10}{\milli\second}, and side peaks appear near $p = \mathrm 2\hbar k_\mathrm{L}$. As $B$ increases, the bifurcation fades. At 750 and \qty{800}{\gauss}, the split is already faint at $t_{\mathrm{shk}} =\qty{6}{\milli\second}$, even though the field reaches \qty{800}{\gauss} only \qty{0.3}{\milli\second} before release. The side peaks weaken in the same way. The bifurcation therefore weakens as the interactions strengthen.

\begin{figure}[ht]
\centering
\includegraphics[width=0.8\linewidth]{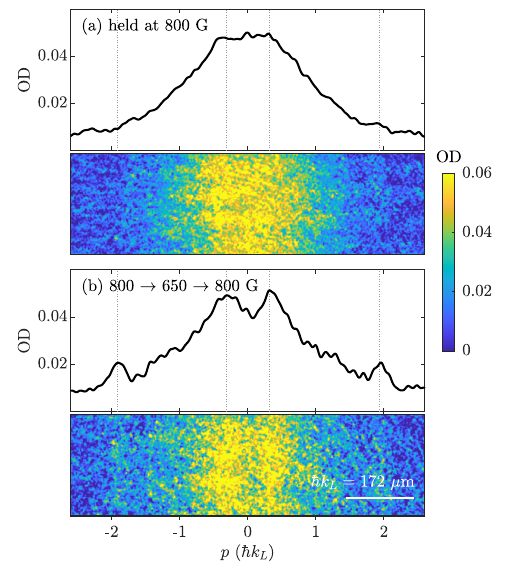}
\caption{Comparison of two different sequences after the condensates are shaken at \qty{800}{\gauss}. In each panel, the upper plot shows the optical density (OD) averaged along the $x$ axis versus momentum $p$ along lattice direction, and the strip below shows the OD image on the same $p$ axis. (a) The field stays at \qty{800}{G} after shaking, and the central peak shows only a faint split. (b) After shaking, the field is ramped to \qty{650}{G} in \qty{1.2}{\milli\second} and then back to \qty{800}{G}. The central peak splits clearly, and side peaks appear near $p = \pm 2\hbar k_\mathrm{L}$. Both clouds were shaken identically, yet the cloud in (b), which is wider across the lattice, shows the stronger split. The dotted lines mark the peaks of (b) in both panels. The scale bar gives $\hbar k_\mathrm{L}$ as a length on the camera after the \qty{7}{\milli\second} TOF, and all images are taken at \qty{800}{G}.}
\label{fig:HFshk}
\end{figure}

% \begin{figure}[ht]
% \centering
% \includegraphics[width=1\linewidth]{fig5_profile.pdf}
% \caption{test figure 5 - showing profile only}
% \label{fig:5 - profile}
% \end{figure}

The fading of bifurcation in \autoref{fig:TOF_evo_Bramp} could be explained by a higher temperature at high field, or the stronger interaction near the Feshbach resonance. To separate these possible effects, we prepare the mBEC at \qty{650}{\gauss}, ramp the field to \qty{800}{\gauss}, and directly shake the lattice at \qty{800}{\gauss} for \qty{0.4}{\milli\second}, where shaking heats up the cloud much more than at 650 G. The results of two field ramping sequences after a high field shaking are compared in \autoref{fig:HFshk}. When the field stays at \qty{800}{\gauss} until imaging, the central peak shows only a faint split in \autoref{fig:HFshk}(a). When we instead ramp the field to \qty{650}{\gauss} in \qty{1.2}{\milli\second}, stay for \qty{0.4}{\milli\second}, and ramp back to \qty{800}{\gauss} for imaging, the central peak splits clearly, and side peaks appeared near $p = \pm 2\hbar k_\mathrm{L}$. However, the cloud in \autoref{fig:HFshk}(b) has a wider vertical ($y$ direction) extent, with a root mean square width of \qty{166}{\mu\meter} versus a width of \qty{149}{\mu\meter} in \autoref{fig:HFshk}(a). Since there is no lattice along this axis, the width of the distribution can be taken as a proxy for temperature. This shows that heat alone does not remove the bifurcation. The main difference arises from the field before release because both clouds expand and imaged at the same field. In \autoref{fig:TOF_evo_Bramp}, the ramp to high field began only after the split had formed, about \qty{0.4}{\milli\second} after shaking at \qty{650}{\gauss}. We thus suggest that the weakening of bifurcation is due to stronger interaction at higher fields before release. One mechanism by which this might occur is the reduction of the effective lattice strength as the molecular size increases. In this picture, each molecule sees the lattice potential averaged over its own size. As the Feshbach resonance is approached, the molecular size increases. For example, at \qty{650}{\gauss} the size is \qty{22}{\nano\meter} and at \qty{800}{\gauss} it is \qty{290}{\nano\meter} \cite{Bartenstein2005PreciseDetermination,Chin2010FeshbachResonances}.

\section{Conclusion}
In conclusion, we extended the cluster lifetime of a bifurcated ${}^6\mathrm{Li}_2$ condensate in a shaken optical lattice double-well band to more than \qty{10}{ms}. Two changes made this possible: a perpendicular trap that confines the dynamics to the lattice axis, and a two-stage shaking sequence that loads the bifurcated clusters stably into the band minima.

The longer lifetime revealed two effects that earlier work could not reach. First, stronger confinement along the lattice put some population at the Brillouin zone edge, and shaking-frequency spectroscopy showed that this population occupies both the ground and first excited bands. This indicates that this population is less likely to originate from relaxation into the first excited band. Instead, it may be caused by scattering from a interference between the primary clusters. Moreover, Bragg differential imaging showed that the shaking sequence controls how strongly the two clusters overlap in real space. Second, ramping the magnetic field toward the Feshbach resonance after the bifurcation has formed leads to the bifurcation disappearing. Further experiments with high field shaking show that a heated cloud still splits clearly when the field returns to \qty{650}{\gauss} before releasing. Therefore, heating can not account for this fading, and interaction strength before releasing decides whether the bifurcation survives. The mechanism of this washout remains open. One candidate is the size of the molecules, which grows by an order of magnitude between \qtylist{650;800}{\gauss} and becomes comparable with the lattice spacing. 

Tuning the interactions adds a control parameter to Floquet-engineered bands. Beyond the resonance, the sequence has the possibility to carry the pairs through the BEC-BCS crossover in an engineered band, where theory predicts exotic pairing such as FFLO phases takes place. The longer cluster lifetime demonstrated here makes such experiments plausible.
% p1: counterdiabatic loading
% p2: t_shk evolution with two perp trap strength. Higher perp trap confinement put more population around zone edge.
% p3: using shaking frequency spectroscopy, we found the zone edge signal roughly splits half between band 0 and 1.
% p4: not likely to be relaxation into the first excited band
% p5: Bragg differential imaging show that overlap could be controlled
% p6: Ramping to high field washes out bifurcation features
% p7: high field shaking reveals might not due to heating
% p8: ramp to 650 before release still split clearly, while a cloud held at 800 G did not

We acknowledge funding from NSF CAREER Award No.~1941985. We also thank C.~A.~R. S\'{a} de Melo for helpful discussions on the experiment.

\bibliography{citing_refs_bibtex}
\end{document}